\documentclass{article}
\usepackage[T1]{fontenc} 
\usepackage[utf8]{inputenc} 
\usepackage{ismir,amsmath,cite,url}
\usepackage{booktabs}
\usepackage{array}
\usepackage{multirow}
\usepackage{graphicx}
\usepackage{color}
\usepackage{enumitem}

\title{Analyzing Pitch Content in Traditional Ghanaian \textit{Seperewa} Songs}

\multauthor
{Kelvin L. Walls$^{1}$ \hspace{1cm} Iran R. Roman$^2$ \hspace{1cm} Kelsey Van Ert$^1$} { \bfseries{Colter Harper$^3$ \hspace{1cm} Leila Adu-Gilmore$^1$}\\
 $^1$ New York University, USA\\
$^2$ Queen Mary University of London, UK\\
$^3$  University at Buffalo, USA\\
}

\def\authorname{KL Walls, IR Roman, K Van Ert, C Harper and L Adu-Gilmore}

\begin{document}

\maketitle
\begin{abstract}

\noindent This study examines the pitch content in traditional Ghanaian \textit{seperewa} (Akan harp-lute) songs, utilizing a unique dataset from field recordings of the mid-twentieth century. We selected 71 songs and used Demucs to isolate vocals from instrumental tracks. We then retrieved the F0 content from these isolated tacks 
and applied Gaussian Mixture Models (GMM) to approximate musical scales. 
Comparative F0 analysis between vocals and \textit{seperewa} revealed higher microtonal deviations from equal temperament in vocal tracks. 
We also note challenges in using MIR tools for musical scale approximation in non-Western music. 
Our research contributes to the quantitative study of pitch in traditional music of Sub-Saharan Africa.

\end{abstract}

\section{Introduction and Motivation}\label{sec:introduction}




This paper examines archived recordings of \textit{seperewa}, a two-course, six or eight string harp-lute that accompanies sung repertoire in Akan languages \cite{nketia1994generative}. The \textit{Akan} are a Ghanaian ethnic and language group that includes the subgroups of \textit{Asante}, \textit{Fante} and \textit{Akuampem} and make up nearly half of the country's population \cite{wilks1989asante}. European documentation has demonstrated its symbolic importance in the Asante Empire in the 18th century \cite{nketia1994generative}. 
Due in part to to the introduction of guitars from Europe, the \textit{seperewa} nearly disappeared by the the mid-20th century, leading to conservation efforts by Ghanaian musicologists Ephraim Amu (1899-1995) and J.H. Kwabena Nketia (1921-2019). 


We chose \textit{seperewa} songs because they demonstrate tuning systems other than equal-temperament (though over time it has been tuned to align with equal-tempered instruments like piano or fretted guitar) \cite{nketia1994generative}. 
The instrument is briefly mentioned in literature addressing West African music; Nketia, however, wrote a comprehensive publication on \textit{seperewa} harmony and melody \cite{agawu2016tonality}. 
A more recent study by McPherson and Obiri-Yeboah examines Akan language encoding in \textit{seperewa} music \cite{mcpherson2023akantone} 
Therefore, due to its cultural significance, studying traditional \textit{seperewa} music could shed light on the original indigenous practices in Africa that made their way to the Americas and shaped the African diasporic musical practice across the globe. 

Musicological research has documented and analyzed musical and social aspects of traditional music across the Africa continent, as well as the influences of Western European religious and military music \cite{agawu2023african}. In particular, there is extensive literature focusing on aspects of rhythm in African music, such as cycle and multidimentionality \cite{locke2011metric, agawu1995african,jacoby2021extreme}, as well as musical connections between Africa and Afro-Latino communities \cite{villepastour2009two, manuel2007mode}. More recent developments have employed computer analysis to look at micro-timing in African drumming to suggest alternative approaches to meter \cite{polak2014timing,sioros2023polyrhythmic}. 
However, there is little research on pitch content in African music that explores microtonal variance and tendencies outside the Western equal-tempered tuning system \cite{collins1989early}. We seek to redress this balance by investigating Ghanaian music’s unique  \emph {scales}. Nketia stated, ``\textit{seperewa} music and Akan songs in general are based on the heptatonic scale, though there remains a great deal of variance from Western tuning systems and tonal logic within this framework'' \cite{nketia1994generative}. 
Therefore, we examine \textit{Akan} pitch through the implied scale and the microtonal content of the \textit{seperewa}'s song repertoire. 
Therefore, our research questions are: \\

\begin{enumerate}[nolistsep]
\item Given a known \textit{seperewa} scale, can we use MIR methods (such as sounrce separation, F0 tracking, and probability density function modeling) to detect its presence in the vocal and instrument pitch content of a \textit{seperewa} song?
\item How ``equal tempered'' are the overall scales we approximate? how microtonally flat or sharp are specific scale degrees? 
\item How similar and different are the scales between the \textit{seperewa} and the vocals?
\end{enumerate}
\textcolor{white}{.}

These questions align with our broader goal of 
decolonizing datasets and studying the effects of music technology’s embedded biases (i.e. the equal-tempered system in synthesis instruments, MIDI protocols, and recording effects like autotune) on traditional and popular musics of the world. In keeping with our decolonizing methods, we obfuscate the original audio material\footnote{the archive granted permission to share some song examples at \texttt{seperewa-pitch-analysis.github.io}} to retain indigenous intellectual property of this archive. Through collaborative efforts with the original authors and their communities, we aim to activate these archives for research in innovative and transformative ways.

\section{Activating An Archival Ghanaian Music Dataset}

This study's dataset is drawn from field recordings collected in the early 1960s by Ghanaian composer and ethnomusicologist Ephraim Amu (1899-1995). These recordings were part of a larger project by Ghanaian ethnomusicologist, composer, and linguist Dr. J.H. Kwabena Nketia (1921-2019), which began in 1952 and aimed to collect recordings of folklore, music, and poetry in, what was then, the United Kingdom’s Gold Coast Colony.








\begin{table*}[ht]
\caption{\textbf{Given a known \textit{seperewa} tuning, which scale components did our scale approximation pipeline (SAP) find?} The third column indicates the number of songs in our corpus where a scale degree (specified by the first two columns) is part of the known \textit{seperewa} tuning. The other columns denote the number of times our SAP "Retrieved" each degree, "Missed" it, or found it although it was "Unexpected" in the \textit{seperewa} tuning. Results are shown for both \textit{seperewa} and vocals. ``\textemdash'' ~denotes that the scale degree was not part of the known \textit{seperewa} tuning in any song.}
\vspace{0.5cm} 
\label{tab:comparison}
\resizebox{\textwidth}{!}{
\begin{tabular}{>{\bfseries}l l c c c c c c c}
\toprule
\toprule
\textbf{Scale} & & \textbf{No. of songs where} & \multicolumn{3}{c}{\textbf{Seperewa}} & \multicolumn{3}{c}{\textbf{Vocals}} \\
\cmidrule(lr){4-6} \cmidrule(lr){7-9}
\textbf{Degree} & \textbf{Quality} & \textbf{in \textit{seperewa} tuning} & \textbf{Retrieved} & \textbf{Missing} & \textbf{Unexpected} & \textbf{Retrieved} & \textbf{Missing} & \textbf{Unexpected} \\
\midrule
\midrule
Tonic & & 71 & 50 & 21 & 0 & 65 & 6 & 0 \\
\cmidrule(lr){1-9}
\multirow{2}{*}{2nd} & Minor & \textemdash & \textemdash & \textemdash & 19 & \textemdash & \textemdash & 14 \\
& Major & 71 & 39 & 32 & 0 & 49 & 22 & 0 \\
\cmidrule(lr){1-9}
\multirow{2}{*}{3rd} & Minor & 3 & 2 & 1 & 17 & 1 & 2 & 26 \\
& Major & 68 & 39 & 29 & 1 & 44 & 24 & 2 \\
\cmidrule(lr){1-9}
4th & & 71 & 47 & 24 & 0 & 55 & 16 & 0 \\
\cmidrule(lr){1-9}
Tritone & & \textemdash & \textemdash & \textemdash & 12 & \textemdash & \textemdash & 11 \\
\cmidrule(lr){1-9}
5th & & 71 & 58 & 13 & 0 & 61 & 10 & 0 \\
\cmidrule(lr){1-9}
\multirow{2}{*}{6th} & Minor & 3 & 2 & 1 & 6 & 2 & 1 & 8 \\
& Major & 68 & 42 & 26 & 1 & 45 & 23 & 1 \\
\cmidrule(lr){1-9}
\multirow{2}{*}{7th} & Minor & \textemdash & \textemdash & \textemdash & 25 & \textemdash & \textemdash & 25 \\
& Major & \textemdash & \textemdash & \textemdash & 44 & \textemdash & \textemdash & 42 \\
\midrule
\midrule
\textbf{Avg. (std.)} & & 53.25 ($\pm$ 29.04) & 34.88 ($\pm$ 19.86) & 18.38 ($\pm$ 11.34) & 10.42 ($\pm$ 13.2) & 40.25 ($\pm$ 23.39) &  13.0 ($\pm$ 8.9) & 10.75 ($\pm$ 13.12) \\
\bottomrule
\bottomrule
\end{tabular}}
\end{table*}

These recordings are the earliest known of this important endangered music tradition, offering an opportunity to examine pitch content, as well as melodic and harmonic structures in music with pre-colonial origins. The importance of these archival materials are also tied to the locations in which they are housed and the individuals that steward those materials. In working with this data set, we are supporting efforts to not only preserve African cultural heritage but also develop resources and institutions to house the data in the countries from which they originated.

\section{Methods}\label{sec:results}

The \textit{seperewa} is traditionally tuned to a heptatonic scale \cite{nketia1994generative}. 
For this analysis, we paired MIR techniques with the informed analysis of 
an expert plucked string instrument performer (he/him), ethnomusicologist, and scholar of traditional Ghanaian music. He provided us with the approximate tuning of the \textit{seperewa} for each song. 
Therefore, for each song we have the \textit{seperewa} heptatonic scale with a ``tonic'' that corresponds to the frequency of an key in an equal-tempered piano (as identified by the \textit{seperewa} expert), and whether the heptatonic scale had major or minor third and sixth degrees. In this scale, the second was always major, the fourth and the fifth were always perfect, and the seventh was always absent.
Note that while the \textit{seperewa} has a known tuning, the vocal parts of these songs can freely sing other notes (such as the tritone or the 7th) or microtonal inflections around all scale degrees. 
Therefore, our analysis does not impose equal-tempered tuning standards as correct. Rather, we analyze the actual values performed by these traditional Ghanaian musicians. 

\subsection{Scale Aproximation Pipeline (SAP).}

\noindent Our pipeline builds upon  other similar ones for F0 vocal analysis \cite{roman2023f0, georgieva2023total, walls2023total}. 
Our archive consists of 71 songs. 
We use Demucs \cite{rouard2023hybrid} to split our songs into isolated vocal and instrumental tracks, then CREPE \cite{kim2018crepe} to extract the instantaneous pitch content (F0). We drop F0 values with a confidence score under 0.8, and convert all values from Hz to cents to interpret results on a linear scale. We also quantize F0 estimates to the nearest tenth (i.e. 10 bins per semitone of 100 cents; each bin encompasses 10 cents) to enhance our analysis of densities.

We determine a track’s scale using the method by Roman et al. \cite{roman2023f0}. This involves training a GMM to identify Gaussian components in the histogram of F0 values for a song. These components signify the notes in a track’s musical scale. 
We limit our analysis to F0 values within a whole step below and an octave above the known \textit{seperewa} tonic. 
After GMM modeling we also recursively average components within 50 cents of each other per song.
To calculate how aligned a scale’s components are with equal temperament, we use the approach described by Roman et al. 
obtaining an $\epsilon_S$ score per song, where a value of zero indicates the highest possible alignment with equal temperament, and 50 complete deviation from it.
Furthermore, we assign ``scale degree'' identities to each component by identifying the tonic based on the known \textit{seperewa} tuning and computing the distance of other components to this tonic anchor. 
For example, if a component is 930 cents above the tonic, that component is labeled as a major sixth with a sharpness of 30 cents.

\section{Results}

\subsection{Finding scales in isolated \textit{seperewa} \& vocal tracks.}

\noindent Table \ref{tab:comparison} quantifies how our SAP identified scale components. 
First, the column ``No. of songs where in \textit{seperewa} tuning'' shows the number of times a scale degree (in the specific quality of major or minor, when appropriate) was known to be part of the \textit{seperewa} tuning.
This knowledge about the \textit{seperewa} tuning was provided by the expert \textit{seperewa} player that we consulted.
Note that the tonic, major second, fourth, and fifth were part of the tuning in all songs.
The 3rd and 6th were predominantly ``major'', and only a handful of songs featured a ``minor'' tuning.  
Note also how no song featured a seventh in the expected \textit{seperewa} tuning, confirming the underlying \textit{seperewa} heptatonic scale according to the \textit{seperewa} expert. Similarly, the minor second and the tritone are never part of the scale. 
The next column, labeled ``Retrieved'', shows that all scale degrees known to be part of the \textit{seperewa} tuning were found in our corpus, although not in all songs. 
For example, the tonic was found in the \textit{seperewa} in only two thirds of the songs, while in the vocals it was found in almost all songs. 
The column labeled as ``missing'' quantifies the number of songs where a scale degree was known to be part of the \textit{seperewa} tuning and was not found by our SAP. 
We also quantified the number of scale degrees found that were ``unexpected'' since they were not part of the \textit{seperewa} tuning. 
Interestingly, in the \textit{seperewa} track (and also in the vocals, although less surprisingly due to the expressive pitch abilities of the voice) our SAP found unexpected minor seconds, minor thirds, and minor sevenths a considerable number of times. 
It also found unexpected tritones and sixth. 
Finally, the bottom row in the table summarizes the average performance of SAP on scale degree retrieval from \textit{seperewa} and vocal tracks, also summarizing the average number of missing and ``unexpected'' components.

Together, these results demonstrate that our SAP can majorly identify the scale degrees in the known \textit{seperewa} tuning in both the isolated \textit{seperewa} and vocal tracks. 
There are limitations to our approach, however. For instance, the number of missed and ``unexpected'' components is not trivial and deserves attention. 
In the case of the SAP applied to the \textit{seperewa}, we hypothesize that the large number of missing components could be caused by the imperfect separation of the instrumental and vocal tracks by Demucs. 
Through manual inspection we found that Demucs allows for the \textit{seperewa} track to leak into the vocal track, particularly in the lower range of the \textit{seperewa}. 
This causes a considerable amount of \textit{seperewa} information to be missing in the instrumental track and could be the underlying factor of the relatively low retrieval of some scale degrees in the \textit{seperewa}. 
In the case of the vocals, the ``unexpected'' scale degrees are easily explained by the voice's ability to freely show microtonal inflections.
The ``unexpected'' components in the \textit{seperewa} can be explained by voice leakage into the instrumental track. 
Future work should look into improving Demucs for the type of recordings we used in this study, and better measuring the challenges of using such model to robustly separate vocal and instrumental tracks.

\subsection{Deviation of scales from equal temperament.}

\begin{figure}[h]
\includegraphics[width=\columnwidth]{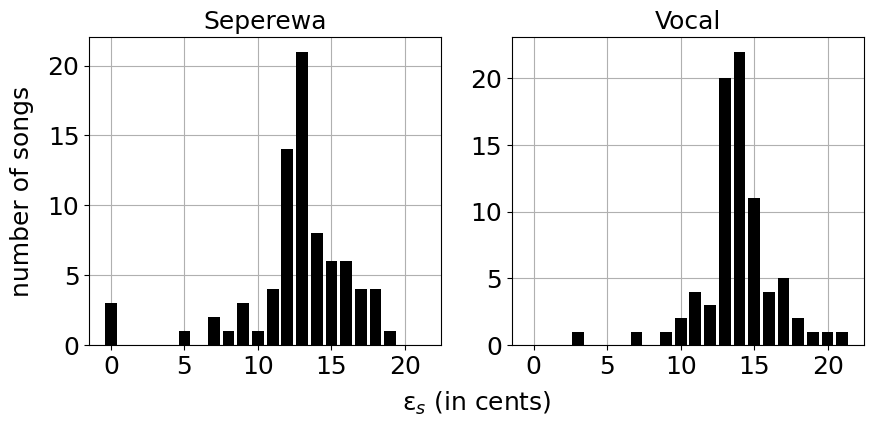}
 \caption{\textbf{How equal-tempered are the scales of songs?} This figure shows the density of the error $\epsilon_S$ (in cents) between the equal-tempered scale and the scale we approximated with our SAP for each song in the corpus. On the x-axis, an $\epsilon_S$ at or close to zero would correspond to an equal-tempered scale, while higher $\epsilon_S$ values denote deviation from equal temperament. Separate histograms show the density of $\epsilon_S$ for the \textit{seperewa} and the vocals.}
  \label{fig:eq_result}
  \vspace{-0.5cm} 
\end{figure}

\noindent After approximating each song's scale in the \textit{seperewa} and vocal tracks, we measured how much each scale deviates from equal temperament by calculating $\epsilon_S$ \cite{roman2023f0}.
Fig. \ref{fig:eq_result} shows the distribution of $\epsilon_S$ values across all songs, separately shown for the \textit{seperewa} and the vocals, with a mean of 11.6 (5.07 std.) and 13.67 (4.92 std.), respectively. 
The larger mean in the vocal $\epsilon_S$ is expected given the voice's ability to freely represent microtonal inflections that deviate from equal temperament.
Also note how the \textit{seperewa} has three songs perfectly aligned with equal temperament ($\epsilon_S=0$), while the same is not observed for the vocals.
In general, these results highlight this music's deviation from equal temperament.

\subsection{Microtonal inflections of scale degrees.}

\begin{figure}
\includegraphics[width=\columnwidth]{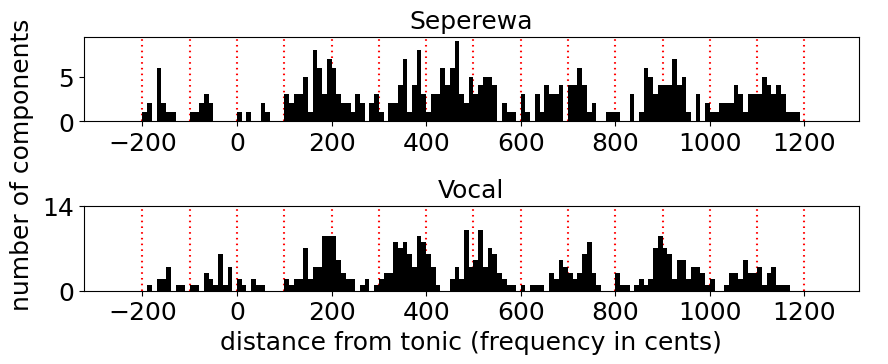}
 \caption{\textbf{How microtonally flat or sharp are individual scale degrees in our corpus with respect to equal temperament?} The density of all scale components found with our SAP on the entire corpus. The vertical red lines correspond to a twelve-tone equal-tempered scale. 
 The x-axis is the frequency distance from the tonic (in cents).
 The top plot shows the component density in the \textit{seperewa} and the bottom is the corresponding plot for vocals.}
  \label{fig:result_degrees}
\end{figure}

We also wanted to understand the microtonality of individual scale degrees. 
Fig. \ref{fig:result_degrees} shows the density of each scale degree in terms of distance in cents from the tonic (recall that 100 cents is one semitone), with red vertical lines referencing the equal tempered twelve-tone scale. 
Results are shown for both the \textit{seperewa} (top) and the vocals (bottom). 

Note the density around 200 cents, corresponding to a microtonally flat ``major second'' in both the \textit{seperewa} and the vocals. 
Similarly, a density between 300 and 400 cents corresponds to a ``third'' that is between minor and major, an effect that is more clearly visible in the vocal scales. 
Next, there are clear densities around 500 and 700, corresponding to the perfect fourth and fifth, respectively.
Finally, both \textit{seperewa} and vocals show a clear density corresponding to a major sixth (900 cents). 
These results give insight into the microtonal tendencies in the individual scale components in this musical corpus. 

\subsection{Comparing the \textit{seperewa} and vocal scales directly.}

\noindent It is also interesting to directly compare the scale components that were common between the \textit{seperewa} and the vocals in a given song. 
This allows to answer whether one is flatter or sharper than the other in general. 
Table \ref{tab:scale_comparison} shows the results of this analysis. 
Here we observe again that major thirds were considerably flatter (negative numbers) in the vocals. 
All other scale degrees showed microtonal deviations between the \textit{seperewa} and the vocals, most times with the vocals being slightly sharper (positive numbers) than the \textit{seperewa}.

\begin{table}[ht!]
\centering
\caption{\textbf{How flat or sharp is each sung scale degree compared to the \textit{seperewa} scale?} Comparison of scale components between \textit{Seperewa} and Vocals as found in SAP. Negative average distances indicate the component was flatter in Vocals than in \textit{Seperewa}, while positive values indicate it was sharper. The third column specifies the number of songs in our corpus where a given  component was found in both the \textit{Sperewa} and the vocals and were used for this analysis. All values are in units of cents.}
\vspace{0.5cm} 
\label{tab:scale_comparison}
\resizebox{\columnwidth}{!}{
\begin{tabular}{>{\bfseries}l l c c}
\toprule
\toprule
\textbf{Scale} & & \textbf{No. of songs with} & \textbf{Avg. Distance} \\
\textbf{Degree} & \textbf{Quality} & \textbf{comp. in both}& \textbf{($\pm$ std.)} \\
\midrule
\midrule
Tonic & & 50 & -2.03 ($\pm$ 36.8) \\
\cmidrule(lr){1-4}
\multirow{2}{*}{2nd} & Minor & 7 & 2.01 ($\pm$ 15.21) \\
& Major & 30 & -1.08 ($\pm$ 36.92) \\
\cmidrule(lr){1-4}
\multirow{2}{*}{3rd} & Minor & 12 & 2.2 ($\pm$ 24.58) \\
& Major & 26 & -12.57 ($\pm$ 42.13) \\
\cmidrule(lr){1-4}
4th & & 37 & 21.42 ($\pm$ 38.65) \\
\cmidrule(lr){1-4}
Tritone & & 4 & 8.67 ($\pm$ 30.01) \\
\cmidrule(lr){1-4}
5th & & 53 & 0.67 ($\pm$ 32.32) \\

\cmidrule(lr){1-4}
\multirow{2}{*}{6th} & Minor & 4 & 17.92 ($\pm$ 32.0) \\
& Major & 28 & 7.08 ($\pm$ 29.11) \\
\cmidrule(lr){1-4}
\multirow{2}{*}{7th} & Minor & 11 & -11.12 ($\pm$ 45.41) \\
& Major & 24 & -21.45 ($\pm$ 43.59) \\
\bottomrule
\bottomrule
\end{tabular}}
\vspace{0.5cm} 
\end{table}

\section{Discussion}

This study analyzed the complex relation between indigenous Ghanaian musical practices and Western tuning influences, revealing systematic and structural deviations and relations to equal temperament.
For example, we identified evidence supporting the existence of a heptatonic scale in the tuning systems and practice of the \textit{seperewa} and the vocal singing that it is usually performed with. 
Our analysis highlights the resilience of traditional tuning practices despite the encroachment of Western musical norms, underscored by the presence of microtonal inflections and the retention of non-equal tempered scales in both \textit{seperewa} and vocals. 
The findings challenge prevailing assumptions about the universality of Western tuning standards and highlight the importance of context-sensitive musicological analysis. 
Future studies should enhance this research by incorporating statistical testing of the preliminary observations made here, necessitating a larger dataset and defined, measurable variables to carry out a statistical analysis that determines the significance of the trends we have described.

\section{Conclusion}

Our research contributes to a nuanced understanding of Ghanaian musical scales, revealing a rich tapestry of sound that defies simple categorization within the Western equal-tempered system \cite{collins1989early,shipley2013living}. By documenting the microtonal variances and tuning discrepancies in \textit{seperewa} songs, this study not only aims at preserving a vital aspect of Ghanaian cultural heritage but also fosters a broader appreciation for the musical diversity that defines the African diaspora. This work underscores the necessity of adopting decolonizing methodologies in musicology, advocating for a more inclusive approach that respects and elevates non-Western musical forms. Future research should continue to explore these themes, further bridging the gaps between traditional African music and its diasporic iterations thereby enriching our global musical heritage.

\bibliography{ISMIRtemplate}

%
%
%
%
%

\end{document}